%% file: combination_manuscript.tex
\newcommand*{\ATLASLATEXPATH}{latex/}
\documentclass[UKenglish,texlive=2011,cernpreprint]{\ATLASLATEXPATH atlasdoc}
\usepackage[subfigure]{\ATLASLATEXPATH atlaspackage}
\usepackage{\ATLASLATEXPATH atlasbiblatex}

\usepackage{\ATLASLATEXPATH atlascontribute}

\usepackage{\ATLASLATEXPATH atlasphysics}
\graphicspath{{logos/}{figures/}}

\usepackage{combination_paper-defs}
\usepackage{multirow}

\input{combination_manuscript-metadata}
\hypersetup{pdftitle={ATLAS draft},pdfauthor={The ATLAS Collaboration}}

\begin{document}

\maketitle

\tableofcontents 

% List of contributors - print here or after the Bibliography.
%\PrintAtlasContribute{0.30}
%\clearpage

%-------------------------------------------------------------------------------
%Introduction
\input{introduction}

%-------------------------------------------------------------------------------
%Atlas detector
\input{atlas_det}
%-------------------------------------------------------------------------------
%Signal and background samples
\input{signal_background_samples}

%-------------------------------------------------------------------------------
%Objects
\input{objects_rec}

%-------------------------------------------------------------------------------
%Event sel (now : Analysis Channels)
\input{event_sel}

%-------------------------------------------------------------------------------
%Combination
\input{combination_proc}
%-------------------------------------------------------------------------------
%Systematics
\input{systematics}

%-------------------------------------------------------------------------------
%Results and conclusion
\input{results}

%-------------------------------------------------------------------------------
%Conclusion
\input{conclusion}
%-------------------------------------------------------------------------------
\section*{Acknowledgements}
%-------------------------------------------------------------------------------
\input{acknowledgements/Acknowledgements}

%-------------------------------------------------------------------------------
%\clearpage
%\appendix
%\part*{Appendix}
%\addcontentsline{toc}{part}{Appendix}
%-------------------------------------------------------------------------------

%-------------------------------------------------------------------------------
% If you use biblatex and either biber or bibtex to process the bibliography
% just say \printbibliography here
\printbibliography

%-------------------------------------------------------------------------------
\newpage

\input{atlas_authlist}

%-------------------------------------------------------------------------------
%-------------------------------------------------------------------------------

\end{document}

%% file: combination_manuscript-metadata.tex
\AtlasTitle{Combination of searches for $WW$, $WZ$, and $ZZ$ resonances in $pp$ collisions at $\sqrt{s} = 8$ \tev\; with the ATLAS detector}

\author{The ATLAS Collaboration}

\AtlasRefCode{EXOT-2014-18}

\PreprintIdNumber{CERN-PH-EP-2015-269}

\AtlasJournal{Phys.\ Lett.\ B.}

\AtlasAbstract{%
The ATLAS experiment at the CERN Large Hadron Collider has performed searches for new, heavy
bosons decaying to $\Wboson\Wboson$, $\Wboson\Zzero$ and $\Zzero\Zzero$ final states in multiple decay
channels using 20.3~\ifb{} of $pp$ collision data at $\sqrt{s} = 8 \tev$.  In the
current study, the results of these searches are combined to provide a
more stringent test of models predicting
heavy resonances with couplings to vector bosons.  Direct searches for
a charged diboson resonance decaying to $\Wboson\Zzero$ in the
\lnull{} ($\ell = \mu, e$) , \llqq{}, \lnuqq{} and
fully hadronic final states are combined
and upper limits on the rate of production times branching ratio to the
 $\Wboson\Zzero$ bosons are compared with
predictions of an extended gauge model with a heavy $W'$
boson.  In addition, direct searches for a neutral diboson resonance
decaying to $\Wboson\Wboson$ and $\Zzero\Zzero$ in the \llqq, \lnuqq, and fully hadronic
final states are combined and upper
limits on the rate of production times branching ratio to the  $\Wboson\Wboson$ and $\Zzero\Zzero$ bosons are compared with predictions for a heavy, spin-2 graviton in an
extended Randall--Sundrum model where the Standard Model fields are allowed to propagate in the bulk of the extra dimension.
} 

%% file: introduction.tex
\section{Introduction}
\label{sec:intro}

The naturalness argument associated with the small mass of the recently discovered Higgs boson~\cite{Aad:2012tfa,Chatrchyan:2012xdj,Aad:2013xqa,Chatrchyan:2012jja} 
suggests that the Standard Model (SM) is conceivably to be extended by a theory that includes additional particles and interactions at the \TeV{} scale.
Many such extensions of the SM, such as extended gauge models~\cite{Altarelli1989,EHLQ1984ref,Quigg2013}, models of warped extra dimensions~\cite{Randall:1999ee, Randall:1999vf, Davoudiasl:2000wi},
technicolour~\cite{Lane:2002sm,Eichten:2007sx,Sannino:2004qp,Belyaev:2008yj}, and more generic composite Higgs models~\cite{Agashe:2004rs, Giudice:2007fh},
predict the existence of massive resonances decaying to pairs of $W$ and $Z$ bosons. 

In the extended gauge model (EGM)~\cite{Altarelli1989} a new, charged vector boson ($W'$) couples to the SM particles. The coupling between the $W'$ and the SM fermions is the same as the coupling between the $W$ boson and the SM fermions. The $W'WZ$ coupling has the same structure as the $WWZ$ coupling in the SM, but is scaled by a factor $c \times (m_W/m_{W'})^2$, where $c$ is a scaling constant, $m_W$ is the $W$ boson mass, and $m_{W'}$ is the $W'$ boson mass. The scaling of the coupling allows the width of the $W'$ boson to increase approximately linearly with $m_{W'}$ at $m_{W'} \gg m_W$ and to remain narrow for $c\sim1$.  For $c=1$ and $m_{W'} > 0.5 \tev $ the $W'$ width is approximately 3.6\% of its mass and the branching ratio of the  $W' \to WZ$ ranges from 1.6\% to 1.2\% depending on $m_{W'}$. Production cross sections in $pp$ collisions at $\sqrt{s} = 8 \tev$ for the $W'$ boson as well as the $W'$ width and branching ratios of $W' \to WZ$ for a selection of $W'$ boson masses in the EGM with scale factor $c=1$ are given in Table~\ref{tab:gravitons_and_wp}.

Searches for a $W'$ boson decaying to $\ell \nu$ have set strong bounds on the mass of the $W'$ when assuming the sequential standard model (SSM)~\cite{ATLAS:2014wra,Khachatryan:2014tva}, which differs from the EGM in that the $W'WZ$ coupling is set to zero. For $c\sim1$ the effect of this coupling on the production cross section of the $W'$  boson at the LHC is very small, thus the production cross section of the $W'$ boson in the SSM and the EGM is very similar. Moreover, due to the small branching ratio of the $W' \to WZ$ in the EGM with the scale factor $c\sim1$, the branching ratios of the $W'$ boson to fermions are approximately the same as in the SSM. Nevertheless, models with narrow vector resonances with suppressed fermionic couplings remain viable extensions to the SM, and thus the EGM provides a useful and simple benchmark in searches for narrow vector resonances decaying to $WZ$.

The ATLAS and CMS collaborations have set exclusion bounds on the production and decay of the EGM $W'$ boson. In searches using the \lnull{} ($\ell \equiv e,\mu$) channel, the ATLAS ~\cite{Aad:2014pha} and CMS~\cite{Khachatryan:2014xja} collaborations have excluded, at the 95\% confidence level (CL),  EGM ($c=1$) $W'$  bosons decaying to $WZ$ for $W'$ masses below 1.52  \tev{} and 1.55 \tev{}, respectively. In addition the ATLAS Collaboration has excluded EGM  ($c=1$) $W'$ bosons for masses below 1.59 \tev{} using the \llqq{}~\cite{Aad:2014xka} channel, and below 1.49 \tev{} using the \lnuqq{}~\cite{Aad:2015ufa} channel. These have also been excluded with masses between 1.3 and 1.5 \tev{} and below 1.7 \tev{} by the ATLAS~\cite{Aad:2015owa} and CMS~\cite{CMSjetref} collaborations, respectively, using the fully hadronic final state. 

Diboson resonances are also predicted in an extension of the original Randall--Sundrum (RS)~\cite{Randall:1999ee, Randall:1999vf, Davoudiasl:2000wi} model with a warped extra dimension. In this extension to the RS model~\cite{ADPS2007,AAS2008,AS2008}, the SM fields are allowed to propagate in the bulk of the extra dimension, avoiding constraints on the original RS model from flavour-changing neutral currents and from electroweak precision measurements. This so-called bulk-RS model is characterized by a dimensionless coupling constant $k/\bar{M}_{\mathrm{Pl}} \sim 1$, where $k$ is the curvature of the warped extra dimension, and $\bar{M}_{\mathrm{Pl}} = M_{\mathrm{Pl}} / \sqrt{ 8 \pi }$ is the reduced Planck mass. In this model a Kaluza--Klein excitation of the spin-2 graviton, \grav, can decay to pairs of $W$ or $Z$ bosons.  For bulk RS models with $k/\bar{M}_{\mathrm{Pl}} = 1$ and for \grav{} masses between 0.5 and 2.5 \tev{}, the branching ratio of \grav{} to $WW$ ranges from 34\% to 16\% and the branching ratio to $ZZ$ ranges from 18\% to 8\%.  The \grav{} width ranges from 3.7\% to 6.2\% depending on the \grav{} mass. Table \ref{tab:gravitons_and_wp} lists widths, branching ratio to $WW$ and $ZZ$ for \grav{}, and production cross sections in $pp$ collisions at 8 \tev{} in these bulk RS models. 

The ATLAS Collaboration has excluded, at the 95\% CL, bulk $\grav \rightarrow ZZ$ with masses below 740 \gev{}, using the \llqq{} channel~\cite{Aad:2014xka}, as well as bulk $\grav \rightarrow WW$  with masses below 760 \gev{}, using the \lnuqq{} channel assuming $k/\bar{M}_{\mathrm{Pl}}=1$~\cite{Aad:2015ufa}. The CMS Collaboration has also excluded at the 95\% CL the $\grav$ of the original RS model, decaying to $WW$ and $ZZ$ with masses below 1.2 \tev{} using the fully hadronic final state~\cite{CMSjetref} and has set limits on the production and decay of generic diboson resonances using a combination of \llqq, \lnuqq{} and fully hadronic final states~\cite{Khachatryan:2014gha}. 

To improve the sensitivity to new diboson resonances, this article presents a combination of four statistically independent searches for diboson resonances previously published by the ATLAS Collaboration~\cite{Aad:2014pha, Aad:2014xka, Aad:2015ufa, Aad:2015owa}. The searches are combined while considering the correlations between systematic uncertainties in the different channels. The first search, sensitive to charged resonances decaying to $WZ$, uses the  \lnull
~\cite{Aad:2014pha} final state. The second search, sensitive to charged resonances decaying to $WZ$ and neutral resonances decaying to $ZZ$, uses  the
 \llqq\;final state~\cite{Aad:2014xka}. The third search, sensitive to charged resonances decaying to $WZ$ and neutral resonances decaying to $WW$, uses the \lnuqq\;final state~\cite{Aad:2015ufa}. Finally, the fourth search, sensitive to  charged resonances decaying to $WZ$ and to neutral resonances decaying to either $WW$ or $ZZ$, uses the fully hadronic final state \cite{Aad:2015owa}. Due to the large momenta of the bosons from the resonance decay, the resonance in this channel is reconstructed with two large-radius jets, and the fully hadronic channel is hereafter referred to as the $JJ$ channel.

To search for a charged diboson resonance decaying to $WZ$ the \lnull, \llqq, \lnuqq, and $JJ$ channels are combined. The result of this combination is interpreted using the EGM $W'$ model with $c=1$ as a benchmark. 

To search for neutral diboson resonances decaying to $WW$ and $ZZ$  the \llqq, \lnuqq, and $JJ$ channels are combined, and the result is interpreted using the bulk \grav{}, assuming $k/\bar{M}_{\mathrm{Pl}}=1$, as a benchmark.

The ATLAS Collaboration has performed additional searches in which new diboson resonances could manifest themselves as excesses over the background expectation. In the analysis presented in Ref.~\cite{Aad:2015kna} the $\ell\ell\ell'\ell'$, $\ell \ell \nu \nu$, $\llqq$ and $\qqnunu$ final states have been explored in the context of the search for a new,  heavy Higgs boson. Also, in the context of searches for dark matter a final state of a hadronically decaying boson and missing transverse momentum~\cite{Aad:2013oja}, and a final state of a leptonically decaying $Z$ boson and missing transverse momentum have been explored~\cite{Aad:2014vka}. These additional searches are not included in this combination. They are not expected to contribute significantly to
the sensitivity of the combined search due to the lower branching ratio in case of the leptonic channels, and the use of only narrow jets in case of the $\qqnunu$ final state.

%% file: atlas_det.tex
\section{ATLAS detector and data sample}
The ATLAS detector is described in detail in Ref.~\cite{atlas-detector}. It covers nearly the entire solid angle~\footnote{ATLAS uses a right-handed coordinate system with its origin at the nominal interaction point (IP) in the centre of the detector and the $z$-axis along the beam pipe. The $x$-axis points from the IP to the centre of the LHC ring, and the $y$-axis points upward. Cylindrical coordinates $(r,\phi)$ are used in the transverse plane, $\phi$ being the azimuthal angle around the beam pipe. The pseudorapidity is defined in terms of the polar angle $\theta$ as $\eta=-\ln\tan(\theta/2)$, and the distance in $(\phi, \eta)$ space as $\Delta R \equiv \sqrt{ (\Delta \phi)^2 + (\Delta \eta )^2}.$} around the interaction point and has an approximately cylindrical geometry. It consists of an inner tracking detector (ID) placed within a 2~T axial magnetic field surrounded by electromagnetic and hadronic calorimeters and followed by a muon spectrometer (MS) with a magnetic field provided by a system of superconducting toroids.  

The results presented in this article use the dataset collected in 2012 by ATLAS from the LHC $pp$ collisions at $\sqrt{s} = 8 \tev$,  using a single-lepton (electron or muon) trigger~\cite{atlas-trigger-2010} with a \pt{} threshold of 24 \gev, or a single large-radius jet trigger with a \pt{} threshold of 360 \gev{}. 
The integrated luminosity of this dataset after requiring data quality criteria to ensure that all detector components have been operational during data taking is 20.3 fb$^{-1}$.
The uncertainty on the integrated luminosity is $\pm 2.8\%$.  It is derived following the methodology detailed in Ref.~\cite{atlas:lumi2011}. 

%% file: signal_background_samples.tex
\section{Signal and background samples}
\label{sec:samples}

%signal
The acceptance and the reconstructed mass spectra for narrow resonances are
estimated with signal samples generated with resonance masses between 
200 and 2500~\GeV, in 100~\GeV{} steps.
The bulk $\grav$ signal events are produced by \calchep3.4~\cite{Belyaev:2012qa} %Pukhov:1999gg
with $k/\bar{M}_{\rm Pl}=1.0$, and the \Wprime{} signal samples are generated with
\pythia8.170~\cite{Sjostrand:2006za}, setting the coupling scale factor $c=1$. 
The factorization and renormalization scales are set to the generated resonance mass. 
The hadronization and fragmentation are modelled with \pythia8 in both 
cases, and the CTEQ6L1~\cite{Pumplin:2002vw} (MSTW2008LO~\cite{Martin:2009iq}) 
parton distribution functions (PDFs) are used for the $\grav$ ($\Wprime$) signal.
The leading-order cross sections and branching ratios for the $W'$ and bulk $\grav$ signal samples 
for selected mass points and assumed values of the coupling parameters are provided in Table~\ref{tab:gravitons_and_wp}. 

\begin{table}[tbp]
\begin{center}
\caption{Leading-order cross sections, widths, and branching ratios for the $W'$ boson in the EGM with scale factor $c=1$ and for the $\grav$ in the bulk RS model with $k/M_{\mathrm{Pl}} = 1$ in $pp$ collisions at $\sqrt{s}=8\TeV$ for a variety of  mass points.}  
\resizebox{\textwidth}{!}{%
\begin{tabular}{c | ccc | cccc}
    \hline
    \hline
     %  & & & \multicolumn{2}{c|}{$W'$} & \multicolumn{3}{c}{$G_{\mathrm{RS}}$} \\
    $m$ & $\Gamma_{W'}$ & $\sigma(W')$ & BR$(W' \rightarrow WZ)$ & $\Gamma_{G_{\mathrm{RS}}}$ & $\sigma(\grav)$ & BR$(\grav \rightarrow WW)$  & BR$(\grav \rightarrow ZZ)$ \\
    $[\tev]$ & $[\gev]$ & $[fb]$ & $[\%]$ & $[\gev]$ & $[fb]$ & $[\%]$ & $[\%]$ \\
    \hline
%    0.5 & 18.0 & 18.4 & 3198.1\phantom{0000} & 1.6 & 3106.0\phantom{000000} & 34 & 18\phantom{0} \\
%0.5 & 18.0 & 18.4 & 200000\phantom{0000} & 1.6 & 3110\phantom{000000} & 34 & 18\phantom{0} \\
0.5 & 18.0 & $2.00 \times 10^5$ & 1.6  & 18.4 &  $3.11 \times 10^{3\phantom{-}}$ & 34 & 18\phantom{0} \\
%    1.0 &  36.0 & 55.4 & 152.2\phantom{000} & 1.3 & 56.0\phantom{0000} & 19 & 10\phantom{0}  \\
%1.0 &  36.0 & 55.4 & 11700\phantom{000} & 1.3 & 56.0\phantom{0000} & 19 & 10\phantom{0}  \\
1.0 &  36.0 & $1.17\times10^4$ & 1.3 & 55.4  & $5.60\times 10^{{1\phantom{-}}}$& 19 & 10\phantom{0}  \\
%    1.5 & 54.0  & 89.5 & 18.7\phantom{00} & 1.3 & 3.14\phantom{00} & 17 & 8 \\
%1.5 & 54.0  & 89.5 & 1440\phantom{00} & 1.3 & 3.14\phantom{00} & 17 & 8 \\
1.5 & 54.0 & $1.44 \times 10^3$ & 1.3  & 89.5  & $3.14 \times 10 ^{0\phantom{-}}$ & 17 & 8 \\
%    2.0 & 73.3 & 122.5\phantom{0} & 3.14 & 1.2  & 0.29\phantom{00} & 16 & 8  \\
2.0 & 73.3 & $2.42 \times 10^2$ & 1.2 & 122.5\phantom{0}  & $2.90 \times 10 ^{-1}$ & 16 & 8  \\
%    2.5 & 90.7 & 155.0\phantom{0} & 0.69 & 1.2 & 0.032\phantom{0} & 16  & 8 \\
2.5 & 90.7 & $5.31 \times 10^{1}$ & 1.2  & 155.0\phantom{0} & $3.20 \times 10 ^{-2}$ & 16  & 8 \\
%    3.0 & 109.0\phantom{0} & 187.2\phantom{0} & 0.19 & 1.2 & 0.0045 & 16 & 8 \\
\hline
\hline
\end{tabular}
}
\label{tab:gravitons_and_wp}
\end{center}
\end{table}

%background
% Z+jets
The backgrounds in the different decay channels are modelled with simulated event samples.
The $\Wjet$ and $\Zjet$ backgrounds are generated using \sherpa 1.4.1~\cite{Gleisberg:2008ta} 
with CT10 PDFs~\cite{Lai:2010vv}.
A separate sample is generated using \alpgen 2.14~\cite{Mangano:2002ea} to
estimate systematic effects, using CTEQ6L1 PDFs and \PYTHIA6~\cite{Sjostrand:2006za} for
fragmentation and hadronization.

The $\Wjet$ and $\Zjet$ 
production cross sections are scaled to next-to-next-to-leading-order 
(NNLO) calculations~\cite{Catani:2009sm}.
% top 
The top quark pair, $s$-channel single-top quark and $Wt$ processes are modelled by the 
\mcatnlo 4.03 generator~\cite{1126-6708-2002-06-029,1126-6708-2003-08-007} with CT10 PDFs, 
interfaced to \HERWIG\cite{Corcella:2000bw} for fragmentation and hadronization and \JIMMY\cite{Butterworth:1996zw} for modelling of the underlying event.
The top quark pair sample is scaled to the production cross section calculated at NNLO in QCD including resummation 
of next-to-next-to-leading logarithmic soft gluon terms with 
Top++2.0~\cite{b-top1,b-top2,b-top3,b-top4,b-top5,b-top6}.
The $t$-channel single-top events are generated by AcerMC~\cite{Kersevan:2004yg} 
with CTEQ6L1 PDFs and \PYTHIA6 for hadronization.
% diboson
The diboson events are produced with the \HERWIG generator and CTEQ6L1 PDFs, except for 
the $\ell \nu \ell' \ell'$ channel which uses \powheg~\cite{Frixione:2007vw,Alioli:2010xd} interfaced to \PYTHIA6.
The diboson production cross sections are normalized to 
next-to-leading-order predictions~\cite{Campbell:1999ah}. Additional diboson samples for the \lnuqq{} channel are produced with the \sherpa generator.  
% QCD
QCD multijet samples are simulated with \PYTHIA6, \HERWIG, 
and \powheg interfaced to \PYTHIA6. 

% general simulation
Generated events are processed with the ATLAS detector simulation program~\cite{Aad:2010ah} based on 
the GEANT4 package~\cite{Agostinelli:2002hh}. Signal and background samples simulated or interfaced 
with \PYTHIA use an ATLAS specific tune of \PYTHIA~\cite{ATLAS:2012uec}.
Effects from additional inelastic $pp$ interactions (pile-up) 
occurring in the same and neighbouring bunch crossings are taken into account 
by overlaying minimum-bias events simulated by \PYTHIA~8.

%% file: objects_rec.tex
\section{Object reconstruction and selection}
\label{sec:obj}
The search channels included in the combination presented in this article use reconstructed electrons, muons, jets and the measurement of the missing transverse momentum.

Electron candidates are selected from energy clusters in the electromagnetic calorimeter within $|\eta| < 2.47$, excluding the transition region between the barrel and the endcap calorimeters ($1.37 < |\eta| < 1.52$),  that match a track reconstructed in the ID. Electrons satisfying `tight' identification criteria are used to reconstruct $W\rightarrow e \nu$ candidates, while $Z \rightarrow ee$ are reconstructed from electrons that satisfy `medium' identification criteria. These criteria are described in Ref.~\cite{Aad:2014fxa}.  Muon candidates are reconstructed within the range $|\eta| < 2.5$ by combining tracks with compatible momentum in the ID and the MS ~\cite{Aad:2014rra}. Only leptons with $\pt > 25 \gev$ are considered.

Backgrounds due to misidentified leptons and non-prompt leptons are suppressed by requiring leptons to be isolated from other activity in the event and also to be consistent with originating from the primary vertex of the event.\footnote{The primary vertex of the event is defined as the reconstructed primary vertex with highest $\sum\pt^2$ where the sum is over the tracks associated with this vertex.} Upper bounds on calorimeter and track isolation discriminants are used to ensure that the leptons are isolated.

Details of the lepton isolation criteria are given in the publications for the  \lnull~\cite{Aad:2014pha}, \llqq~\cite{Aad:2014xka}, and \lnuqq~\cite{Aad:2015ufa} channels.

Jets are formed by combining topological clusters reconstructed in the calorimeter system~\cite{topoclusters}, which are calibrated in energy with the local calibration weighting scheme~\cite{lcw} and are considered massless. The measured energies are corrected for losses in passive material, the non-compensating response of the calorimeters and pile-up~\cite{Aad:2011he}

Hadronically decaying vector bosons with low \pt{} ($\lesssim 450 \gev{}$) are reconstructed using a pair of jets. The jets are formed with the anti-$k_t$ algorithm~\cite{antikt} with a radius parameter $R=0.4$. These jets are  hereafter referred to as small-$R$ jets. Only small-$R$ jets with $|\eta| < 2.8$ $(2.1)$ and \pt{} > 30 \gev{} are considered for the \lnuqq{} (\llqq) channel. For small-$R$ jets with \pt{} < 50 \gev{} it is required that the summed scalar \pt{} of the tracks matched to the primary vertex accounts for at least 50\% of the scalar summed \pt{} of all tracks matched to the jet. Jets containing hadrons from $b$-quarks are identified using a multivariate $b$-tagging algorithm as described in Ref.~\cite{btagging}.

Hadronically decaying vector bosons with high \pt{} ($\gtrsim 400 \gev{}$) 
can be reconstructed as a single jet with a large radius parameter, or large-$R$ jet, due to the collimated nature of their decay products. These large-$R$ jets, hereafter denoted by $J$, are first formed with the Cambridge--Aachen (C/A) algorithm ~\cite{Dokshitzer:1997,cambridgeaachen1} with a radius parameter $R=1.2$. After the jet formation a set of criteria is applied to identify the jet 
as originating from a hadronically decaying boson (boson tagging). A grooming algorithm is applied to the jets to reduce the effect of pile-up and underlying event activity and to identify a pair of subjets associated with the quarks emerging from the vector boson decay. The grooming algorithm, a variant of the mass-drop filtering technique~\cite{bdrs}, is described in detail in Ref.~\cite{Aad:2015owa}. The grooming procedure provides a small degree of discriminating power between jets from hadronically decaying bosons and those originating from background processes.

Jet discrimination is further improved by imposing additional requirements on the large-$R$ jet properties. First, in all of the channels using large-$R$ jets, a requirement on  the subjet momentum-balance found at the stopping point of the grooming algorithm, $\sqrt{y} > 0.45$, \footnote{$\sqrt{y} \equiv \text{min}({\pt}_{j_1},{\pt}_{j_2}) \frac{\Delta R_{(j_1,j_2)}}{m_0}$, where $m_0$ is the mass of the groomed jet at the stopping point of the splitting stage of the grooming algorithm, ${\pt}_{j_1}$ and ${\pt}_{j_2}$ are the transverse momenta of the subjets at the stopping point of the splitting stage of the grooming algorithm and $\Delta R_{\left( j_1,j_2 \right)}$ is the distance in ($\phi,\eta$) space between these subjets.} is applied to the jet. Second, jets are required to have the groomed jet mass within a selection window. Due to the different backgrounds affecting each of the search channels, different mass windows are used for each channel. In the single lepton and dilepton channels, mass windows of $65 < m_{J} < 105 \gev$ and $ 70 < m_{J} < 110 \gev$, where $m_{J}$ represents the jet mass, are used for selecting $W$ and $Z$ bosons. In the fully hadronic channel, mass windows of $69.4 < m_{J} < 95.4 \gev$ and $79.8 < m_{J} < 105.8 \gev$, which are $\pm 13 \gev$ around the expected $W$ or $Z$ reconstructed mass peak,  are used for selecting $W$ or $Z$ boson candidates respectively. 

The high-\pt{} jets in background events are expected to have a larger charged-particle track multiplicity than the jets emerging from boson decays. This is due to the higher energy scale involved in the fragmentation process of background jets and also due to the larger color charge of gluons in comparison to quarks. Hence, to improve the sensitivity of the search in the fully hadronic channel, a requirement on the charged-particle track multiplicity matched to the large-$R$ jet prior to the grooming, $n_{\mathrm{trk}} < 30$,   is used to discriminate between jets originating from boson decays and jets from background processes. Charged-particle tracks reconstructed with the ID and consistent with particles originating from the primary vertex and with \pt{} $\geq 500$ \mev{} are matched to a large-$R$ jet by representing each track by a ``ghost'' constituent that is collinear with the track at the perigee with negligible energy during jet formation~\cite{Cacciari:2008gn}. 

The missing transverse momentum $\met$ is calculated from the negative vector sum of the transverse momenta of
all reconstructed objects, including electrons, muons, photons and jets, as well as calibrated energy 
deposits in the calorimeter that are not associated to these objects, as described in Ref.~\cite{Aad:2012re}.

%% file: event_sel.tex
\section{Analysis channels}
\label{sec:evsel}
%Introduction
The selections in the four analysis channels \lnull, \llqq, \lnuqq{} and 
$JJ$ are mutually exclusive and therefore the channels are statistically independent. 
This independence is enforced by the required lepton multiplicity of the events at a pre-selection stage, 
with lepton selection criteria looser than those finally applied in the individual channels.
%The statistical independence
 %of the different search channels is enforced based on the lepton multiplicity of the events at a pre-selection stage. 
%The lepton selection criteria are looser than the final lepton selection criteria applied in the individual channels. 
The searches in the individual channels are described in detail in their corresponding publications~\cite{Aad:2014pha, Aad:2014xka, Aad:2015ufa, Aad:2015owa}.
Table~\ref{tab:backgrounds}
summarises the dominant backgrounds affecting each of the individual channels and the methods used to estimate these backgrounds. 
Summaries of the event selection and classification criteria are given in Tables~\ref{tab:selectionsummary} and~\ref{tab:classificationsummary}. 

\begin{table}[!htbp]
\centering
\caption{Dominant background to the individual channels and their estimation methods.}
\label{tab:backgrounds}
\begin{tabular}{l|ll}
\hline
Channel                                & Dominant background           & Estimation method          \\ \hline \hline                       
\multirow{1}{*}{$\lnull$} & $WZ$ production               & MC (\powheg)    \\ \hline                             
\multirow{1}{*}{$\llqq$}        & $Z$+jets                      & MC (\sherpa), normalisation and shape correction data driven \\ \hline                              
\multirow{1}{*}{$\lnuqq$}        & $W/Z$+jets                    & MC (\sherpa), normalisation and shape correction data driven \\ \hline      
\multirow{1}{*}{$JJ$}                  & QCD jets                      & data driven     \\ \hline                                               
\end{tabular}
\end{table}

%lvll
The \lnull{} analysis channel is described in detail in Ref.~\cite{Aad:2014pha}. 
For the purpose of combination the binning of the diboson candidates' invariant mass distribution is adjusted.
The \lnull{} channel requires exactly three leptons with $\pt>25\GeV$, 
of which at least one must be geometrically
matched to a lepton reconstructed by a trigger algorithm. Events with additional 
leptons with $\pt>20\GeV$ are vetoed. At least one pair of oppositely-charged, same-flavour leptons is required to have
an invariant mass within the $\Zzero$ mass window $|m_{\ell \ell}-m_{\Zzero}| < 20\GeV$. If there are two acceptable combinations satisfying this 
requirement the combination with the mass value closer to the $\Zzero$ boson mass is chosen as the $\Zzero$ candidate. 
The event is required  to have $\MET>25\GeV$. The $\Wboson$ candidate is reconstructed from the third
lepton, assuming the neutrino is the only source of $\MET$ and constraining the $(\ell^{3rd},\MET)$ system
to have the pole mass of the $\Wboson$. This constraint results in a quadratic equation with two solutions for 
the longitudinal momentum of the neutrino. If the solutions are real, the one with the smaller absolute value is used. If the
solutions are complex, the real part is used.
%meaning pole mass?
% low mass and high mass regions
To enhance the signal sensitivity, the rapidity difference must satisfy $\Delta y(\Wboson,\Zzero) < 1.5$ and 
requirements are placed on the azimuthal angle difference $\Delta \phi(\ell^{3rd},\MET)$. Exclusive
{\it high-mass} and {\it low-mass} regions are defined with $\Delta \phi(\ell^{3rd},\MET)<1.5$ for boosted \Wboson{} bosons  
and $\Delta \phi(\ell^{3rd},\MET)>1.5$ for  \Wboson{} bosons at low $\pt$, respectively.
%lvll
The main background sources in the \lnull{}  channel are SM $\Wboson\Zzero$ and $\Zzero\Zzero$ processes with leptonic decays of the $\Wboson$ and $\Zzero$ bosons, and
are estimated from simulation. 
Other background sources are $\Wboson/\Zzero+$jets, top quark and multijet production, 
where one or several jets are mis-reconstructed as leptons. 
To estimate these backgrounds the mis-reconstruction rate of jets as leptons is determined with data-driven methods,
and applied to control data samples with leptons and one or more jets. 

%lljj
The \llqq{} analysis channel is described in detail in Ref.~\cite{Aad:2014xka}. 
%and no changes 
%are applied to this channel in the combination with the other channels with respect to this publication.
The \llqq{} channel requires exactly two leptons, 
having the same flavour and with $\pt>25\GeV$.
Muon pairs are required to have opposite charge.
At least one lepton is required to be  
matched to a lepton reconstructed by a trigger algorithm. The invariant mass of the 
lepton pair must be within $25\GeV$ of the $\Zzero$ mass. 
Three regions ({\it merged, high-}\pt{} {\it resolved} and {\it low-}\pt{} {\it resolved})
are defined to optimize the selection for different mass ranges.
The merged region requirements are $\pt(\ell\ell) > 400\GeV$ and a groomed large-$R$ jet
described in Section~\ref{sec:obj} 
with $\pt(J) > 400\GeV$ and satisfying the boson-tagging criteria.
The high-$\pt$ resolved region is defined by $\pt(\ell\ell) > 250\GeV$, $\pt(jj) > 250\GeV$, 
and the low-$\pt$ resolved region requires $\pt(\ell\ell) > 100\GeV$, $\pt(jj) > 100\GeV$.
The invariant mass requirement on the jet system is $70\GeV < m_{jj/J} < 110\GeV$.
The three regions are made exclusive by applying the above selections in sequence, starting with the merged region, and  progressing with
the high-$\pt$ and then the low-$\pt$ resolved regions.  
%lljj
The main background sources in the \llqq\, channel are $\Zzero + $jets, followed by top-quark pair and non-resonant vector-boson pair production.
Background estimates are based on simulation. 
Additionally, for the main background source, 
$\Zzero + $jets, the shape of the invariant mass distribution is modelled with simulation, 
while the normalization and a linear shape correction are determined from data in a control
region, defined as the side-bands of the $q\bar{q}$ invariant mass distribution 
outside the signal region.

%lvjj
The \lnuqq{} analysis channel is described in detail in Ref.~\cite{Aad:2015ufa}.
% and no changes 
%are applied to this channel in the combination with the other channels with respect to this publication. 
In the \lnuqq{} channel exactly one lepton with $\pt>25\GeV$ and matched to a lepton reconstructed
by the trigger is required.
The missing transverse momentum in the event is required to be $\MET>30\GeV$. Similar to the \llqq{} channel 
the event selection contains three different mass regions of the signal, 
referred to as {\it merged, high-}$\pt$ {\it resolved} and {\it low-}$\pt$ {\it resolved} regions. 
In the merged region where the hadronic decay products merge into a single jet,
a groomed large-$R$ jet with $\pt>400\GeV$ and $65\GeV < m_{J} < 105 \GeV$ is required. 
The leptonically decaying $\Wboson$ candidate is reconstructed using the 
same $\Wboson$ mass constraint technique used in the \lnull{} channel. The leptonically decaying $\Wboson\to\ell\nu$ 
must have $\pt(\ell\nu) > 400\GeV$, where $\pt(\ell\nu)$ is reconstructed from 
the sum of the charged-lepton momentum vector and the $\MET$ vector.
To suppress the background from 
top-quark production, events with an identified $b$-jet separated by $\Delta R > 0.8$ from the large-$R$ jet are rejected. Additionally, in the electron channel
the leading large-$R$ jet and $\MET$ are required to be separated by $\Delta \phi (\MET,J) > 1$
to reject multi-jet background.
If the event does not satisfy the criteria of the merged region, 
the resolved region selection criteria are applied. In the high-$\pt$ resolved region, 
two small-$R$ jets with 
$\pt>80\GeV$ are required to form the hadronically decaying $\Wboson$/$\Zzero$
candidate with a transverse momentum of $\pt(jj) > 300\GeV$ and an invariant mass
of $65\GeV < m_{jj} < 105 \GeV$. The leptonically decaying 
$\Wboson\to\ell\nu$ must have $\pt(\ell\nu) > 300\GeV$. The event is rejected if a $b$-jet is
identified in addition to the two leading jets. In the electron channel
the leading small-$R$ jet and $\MET$ are required to be separated by 
$\Delta \phi (\MET,j) > 1$. If the event does not pass the selection requirements of the high-$\pt$ resolved region
the selection of the low-$\pt$ resolved region is used, where $\pt(jj)>100\GeV$ and $\pt(\ell\nu)>100\GeV$ are applied.
%lvjj
The dominant background in the \lnuqq{} channel is $\Wboson/\Zzero+$jets production,
followed by top quark production, and multijet and diboson processes.
The shape of the invariant mass distribution for the $\Wboson/\Zzero+$jets background  
is modelled by simulation, 
while the normalization is determined from data in a control
region, defined as the side-bands of the $q\bar{q}$ invariant mass distribution 
outside the signal region. The $\pt(\Wboson)$ distribution of the $\Wjet$ simulation 
is corrected using data to improve the modelling.
The sub-dominant background processes are estimated using simulation only (diboson),
or simulation and data-driven techniques (multijet, top quark).

%JJ
The $JJ$ analysis channel is described in detail in Ref.~\cite{Aad:2015owa}.
For the combined \grav{} search the analysis is extended, combining the $WW$ and $ZZ$ 
selections into a single inclusive analysis of both decay modes.
The analysis of the fully hadronic decay mode selects events that pass a large-$R$ jet trigger\footnote{The trigger uses anti-$k_t$ jets with $R=1.0$.}
with a nominal threshold of $360\GeV$ in transverse momentum and have at least 
two large-$R$ jets within $|\eta|<2.0$, a rapidity difference between the two jets of $|\Delta y_{12}| < 1.2$, 
and an invariant mass of the two jets of $m(JJ) > 1.05\TeV$. 
Events that contain one or more leptons
with $\pt>20\GeV$ or missing transverse momentum in excess of $350\GeV$ are vetoed.
The large-$R$ jets must satisfy the boson-tagging criteria 
described in Section~\ref{sec:obj}. 
Furthermore, the dijet $\pt$ asymmetry defined as 
$A=({\pt}_1 - {\pt}_2) / ({\pt}_1 + {\pt}_2)$ must be less then 0.15 to avoid mis-measured jets. 
In the search for the EGM $W'$ decaying to $WZ$, events are selected by requiring one $W$
boson candidate and one $Z$ boson candidate in each event by applying the selections described in
Section~\ref{sec:obj}. In the search for the bulk \grav{} decaying to $WW$ and $ZZ$, events 
are selected by requiring two $W$ boson or two $Z$ boson candidates by applying the selections described
in Section ~\ref{sec:obj}. Due to the overlapping jet mass windows applied to select $W$ and $Z$ candidates, 
the selection for the EGM $W'$ and the bulk \grav{} are not exclusive and about 20\% of the inclusive
event sample is shared. 
%JJ
In the fully hadronic channel the dominant background is dijet production. 
The dijet background is estimated by a parametric fit
with a smoothly falling function to the observed dijet mass spectrum in the data. 
Only diboson resonances with mass values $>1.3$ \TeV{} are considered as signal for this analysis channel.

%The fit accurately models the steeply falling QCD dijet mass spectrum.

\begin{figure}[th!]
  \begin{center}
     \subfigure[]
              {
     \includegraphics[width=0.485\textwidth]{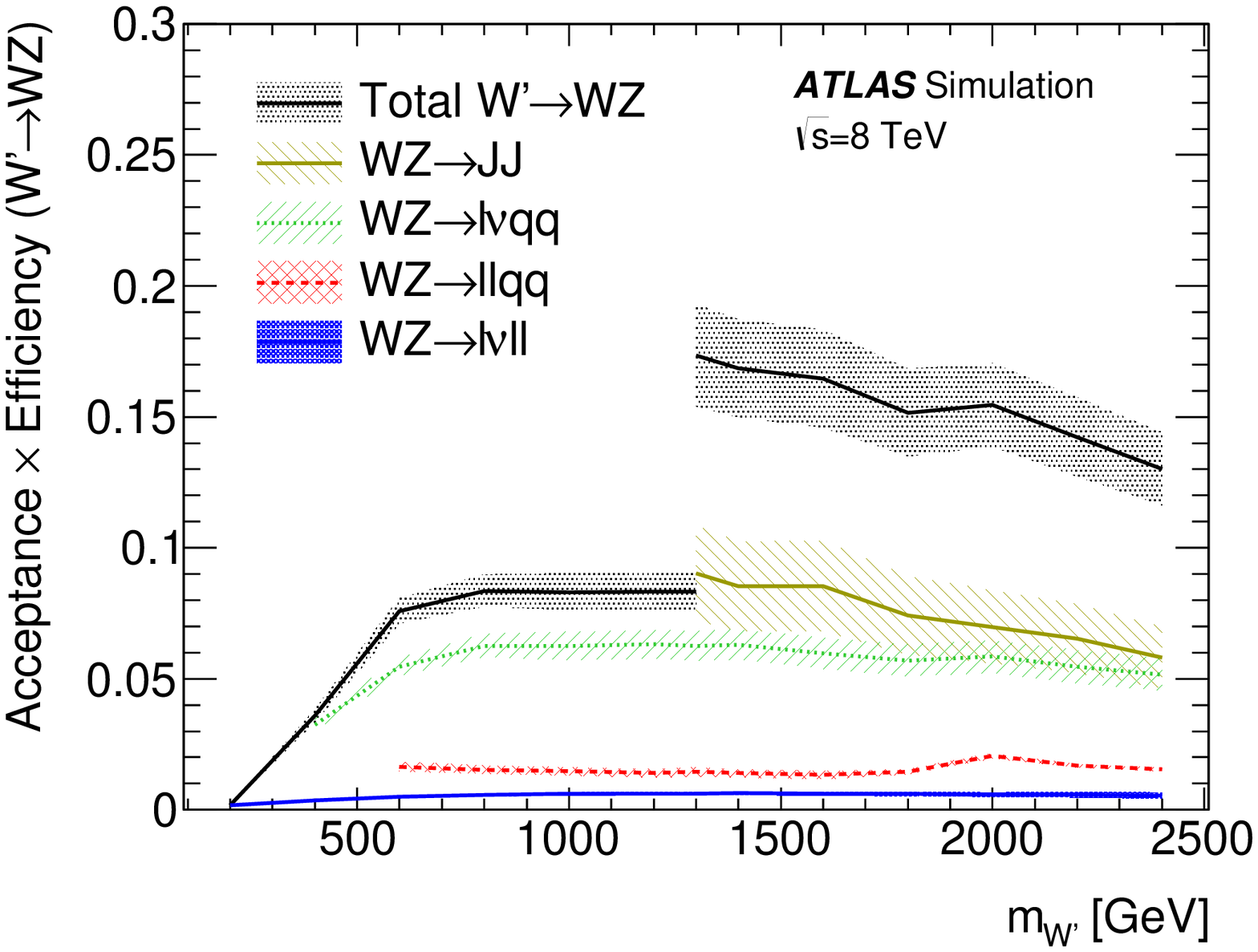}\label{fig:wprime_acceff_summary}
              }
    \subfigure[]
              {
     \includegraphics[width=0.485\textwidth]{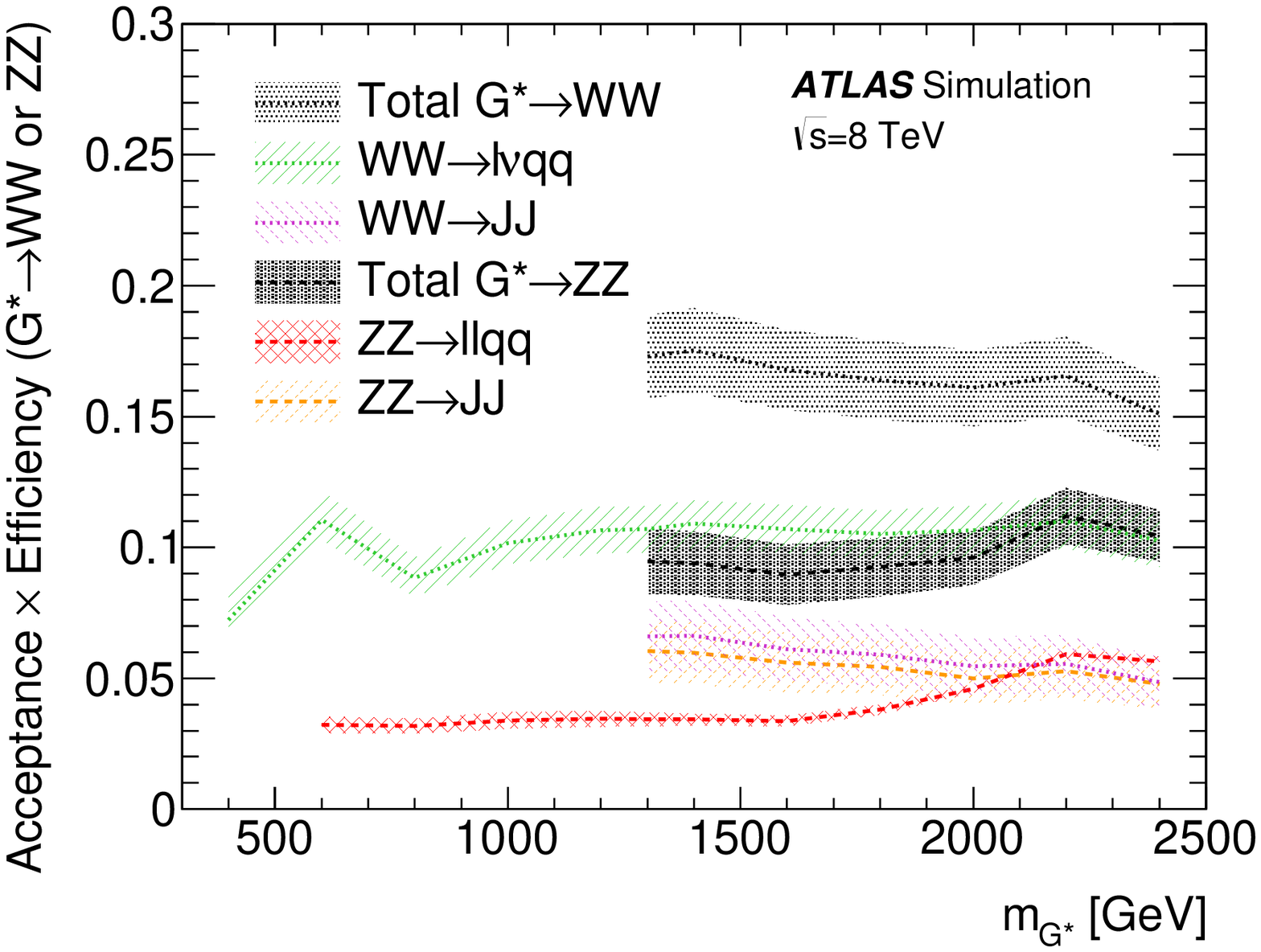}\label{fig:Grav_acceff_summary}
              }
   \end{center}
\vspace{-20 pt}
     \caption{
     Signal acceptance times efficiency for the different analyses entering the combination for (a) the EGM $W'$ model and (b) for the bulk \grav{} model. The branching ratio of the new resonance to dibosons is included in the denominator. 
      The error bands represent the combined statistical and systematic uncertainties. }
\label{fig:accteff}
\end{figure}

%Acceptance
The selections described above have a combined acceptance times efficiency of up 
to 17\% for $\grav\to WW$, up to 11\% for $\grav\to ZZ$, and up to 17\% for $W'\to WZ$.  
The acceptance times efficiency includes the $\Wboson$ and $\Zzero$ branching ratios.
Figs.~\ref{fig:wprime_acceff_summary} and \ref{fig:Grav_acceff_summary} 
summarize the acceptance times efficiency for the different analyses 
as a function of the $W'$ mass and of the $\grav$ mass, considering 
only decays of the resonance into $VV$, where $V$ denotes a $W$ or a $Z$ boson.

\begin{table}[]
\centering
\caption{
  Summary of the event selection requirements in the different search channels. The selected events are further classified into different kinematic categories as listed in Table~\ref{tab:classificationsummary}.
}
\label{tab:selectionsummary}
% STRETCH
\renewcommand{\arraystretch}{1.2} 
\resizebox{\textwidth}{!}{
\begin{tabular}{l|l|l|l|l}
\hline
\hline
\multirow{2}{*}{Channel}               & \multirow{2}{*}{Leptons}                 & \multirow{2}{*}{Jets}                                       & \multirow{2}{*}{\MET}           & \multirow{2}{*}{Boson identification}            \\ 
                                       &                                          &                                                             &                                &                                                  \\ \hline
\multirow{2}{*}{\lnull} 			   & 3 leptons			      				  & \multirow{2}{*}{--}                                           & \multirow{2}{*}{$\MET>25\GeV$} & \multirow{2}{*}{$|m_{\ell \ell}-m_{\Zzero}| < 20\GeV$} 		\\ 
                                       & $\pt>25\GeV$                             &                                                             &                                & 													\\ \hline
\multirow{3}{*}{\llqq}        		   & 2 leptons							      & \multirow{1}{*}{2 small-$R$ jets or 1 large-$R$ jet}        & \multirow{3}{*}{--}            & $|m_{\ell \ell}-m_{\Zzero}| < 25\GeV$           \\  
                                       & $\pt>25\GeV$                             &  $\pt>30\GeV$                                               &                  				 & $70\GeV < m_{jj} < 110\GeV$        				\\ 
                                       &                                          &                                                             &                                & $70\GeV < m_J < 110\GeV$ , $\sqrt{y} > 0.45$     \\ \hline
\multirow{3}{*}{\lnuqq}        		   & 1 lepton   							  & {2 small-$R$ jets or 1 large-$R$ jet}                       						 & \multirow{2}{*}{$\MET>30\GeV$} & $65\GeV < m_{jj} < 105\GeV$ \\  
                                       & \multirow{1}{*}{$\pt>25\GeV$}            & $\pt>30\GeV$											    &                                & $65\GeV < m_J < 105\GeV$, $\sqrt{y} > 0.45$      \\  
									   &										  & {No $b$-jet} with $\Delta R (b, \Wboson/\Zzero)> 0.8$       &								 &													\\ \hline
\multirow{2}{*}{$JJ$}                  & \multirow{2}{*}{lepton veto}                & \multirow{2}{*}{2 large-$R$ jets, $|\eta|<2.0$, $\pt>540\GeV$}& \multirow{2}{*}{$\MET<350\GeV$}& $|m_{W/Z} - m_{J}| < 13\GeV$                   \\ 
            				           &                                          &                   				                         	& 				                 &$\sqrt{y} > 0.45$,  $n_{\mathrm{trk}} < 30$       \\ 
\hline \hline
\end{tabular}
}

\end{table}

% SECOND TABLE

\begin{table}[!htbp]
\centering
\caption{
  Summary of the event classification requirements in the different search channels. The classifications are 
  mutually exclusive, applying the requirements in sequence beginning with the high-\pt{} merged, followed by 
  the high-\pt{} resolved and finally with the low-\pt{} resolved classification. 
}
\label{tab:classificationsummary}
% STRETCH
\renewcommand{\arraystretch}{1.35} 
\resizebox{\textwidth}{!}{
\begin{tabular}{l|l|l|l}
\hline
\hline

\multirow{1}{*}{Channel}               &  High-$\pt$ merged                        & High-$\pt$ resolved (high mass)    & Low-$\pt$ resolved (low mass)           \\ \hline
\multirow{2}{*}{\lnull}  			   & \multirow{2}{*}{--} 					   & \multicolumn{2}{c}{$\Delta y(\Wboson,\Zzero) < 1.5$}                         \\ \cline{3-4} 
                                       &                                           &  $\Delta \phi(\ell^{3rd},\MET)<1.5$ & $\Delta \phi(\ell^{3rd},\MET)>1.5$    \\ \hline
\multirow{2}{*}{\llqq}        		   & $\pt(\ell\ell) > 400\GeV$                 & $\pt(\ell\ell) > 250\GeV$          	       & $\pt(\ell\ell) > 100\GeV$          \\ 
                                       & $\pt(J) > 400\GeV$                        & $\pt(jj) > 250\GeV$                 & $\pt(jj) > 100\GeV$                \\ \hline
\multirow{4}{*}{\lnuqq}        		   & {1 large-$R$ jet, $\pt>400\GeV$} 		   & {2 small-$R$ jets, $\pt>80\GeV$}   & {2 small-$R$ jets, $\pt>30\GeV$}   \\ \cline{3-4} 
                                       &\multirow{2}{*} {$\pt(\ell\nu) > 400\GeV$} & $\pt(jj) > 300\GeV$                & $\pt(jj) > 100\GeV$                \\ 
                                       &                                           & $\pt(\ell\nu) > 300\GeV$           & $\pt(\ell\nu) > 100\GeV$           \\ \cline{2-4}
                                       & \multicolumn{3}{c}{$\Delta \phi (\MET,j) > 1$ (electron channel)} 			          \\  \hline
\multirow{2}{*}{$JJ$}                  & {$|\Delta y_{12}| < 1.2$} 				   &     \multicolumn{2}{l}{\multirow{2}{*}{--}}                             \\  
                                       & {$m(JJ) > 1.05\TeV$}      				   &    \multicolumn{2}{l}{}                                                 \\ \hline \hline

\end{tabular}
}

\end{table}

%%%

%% file: combination_proc.tex
 \section{Statistical procedure}
 \label{sec:combproc}
 
 The combination of the individual channels  proceeds with a simultaneous analysis of the invariant mass distributions of the
 diboson candidates in the different channels. For each hypothesis being tested, only the channels sensitive to that hypothesis 
 are included in the combination. 
The signal strength, $\mu$, defined as a scale factor on the cross section times branching ratio predicted by the signal hypothesis,
  is the parameter of interest.  The analysis follows the Frequentist approach with a test statistic based on the profile-likelihood ratio~\cite{Cowan:2010js}. The test statistic extracts information on the 
  signal strength from a binned maximum-likelihood fit of the signal-plus-background  model to the data. 
  The effect of a systematic uncertainty $k$ on the likelihood is modelled with a nuisance parameter, $\theta_{k}$, constrained with a corresponding probability 
 density function $f(\theta_k)$, as explained in the publications corresponding to the individual channels~\cite{Aad:2014pha,Aad:2014xka,Aad:2015ufa,Aad:2015owa}.  In this manner, correlated effects across the different channels are modelled by the use of a common nuisance parameter 
 and its corresponding probability 
 density function.  The likelihood model, $ \mathcal{L} $,  is given by:

 \begin{equation}
 \mathcal{L} = \prod_{c} \prod_{i} \mathrm{Pois}\left(n^{\mathrm{obs}}_{i_c} \left| n^\mathrm{sig}_{i_c}(\mu, \theta_k) + n^{\mathrm{bkg}}_{i_c}( \theta_k) \right. \right) \prod_{k} f_{k}(\theta_k)
 \end{equation} 
 where the index $c$ represents the analysis channel, and $i$ represents the bin in the invariant mass distribution, $n^{\mathrm{obs}}$, the observed number of events, $n^\mathrm{sig}$ the number of expected signal events, and $n^\mathrm{bkg}$ the expected number of background events.

 The compatibility between the observations of different channels with a common signal strength of a particular resonance model and mass is quantified using a profile-likelihood-ratio test. The corresponding profile-likelihood ratio is

 \begin{equation}
 \lambda({\mu}) = \frac{  \mathcal{L} \left( {\mu}, \mathbf{\hat{\hat{\theta}}}\left( {\mu} \right) \right) } {  \mathcal{L} \left( \hat{\mu}_A, \hat{\mu}_B, \mathbf{{\hat{\theta}}} \right)}, 
 \end{equation}

where $\mu$ is the common signal strength, $\hat{\mu}_A$ and $\hat{\mu}_B$ are the unconditional maximum likelihood (ML) estimators of the independent signal strengths in the channels being compared, 
$\hat{\mathbf{\theta}}$ are the unconditional ML estimators for the nuisance parameters, and $\hat{\hat{\mathbf{\theta}}}(\mu)$ are the conditional ML estimators of $\mathbf{\theta}$ for a given value of $\mu$.
The compatibility between the observations is tested by the probability of observing $\lambda(\hat{\mu})$, where $\hat{\mu}$ is the ML estimator 
for the common signal strength for the model in question. If the two channels being compared have a common signal strength, i.e. $\mu = \mu_A = \mu_B$, then 
in the asymptotic limit $-2 \log (\lambda(\hat\mu))$ is expected to be $\chi^2$ distributed with one degree of freedom.  

The significance of observed excesses over the background-only
 prediction is quantified using the local $p$-value ($p_0$), defined as the probability of the background-only model to produce a signal-like fluctuation at least as large as observed in the data. 
Upper limits on $\mu$ for $W'$ in the EGM and \grav{} in the bulk RS model at the simulated resonance masses are evaluated at the 95\% CL following the $CL_{s}$ prescription~\cite{Read:2002hq}.
Lower mass limits at the 95\% CL for new diboson resonances in these models are obtained by finding the maximum resonance mass where the 95\% CL upper limit on $\mu$ is less or equal to $1$. This mass is found by interpolating  between the limits on $\mu$ at the simulated signal masses. The interpolation assumes monotonic and smooth behaviour of the efficiencies for
the signal and background processes, and that the impact of the variation of signal mass distributions between adjacent test masses is negligible.
 
 In the combined analysis to search for $W'$ resonances, all four individual channels 
 are used. For the charge-neutral bulk \grav, only the $\ell\nu qq$, $\ell\ell qq$, and the $JJ$ channels
 contribute to the combination, and in the case of the fully hadronic channel, a merged signal region resulting from the union of the $WW$ and $ZZ$ signal regions is used in the analysis. 
 The background to this merged signal region is estimated using the same technique as for the 
 individual signal regions. Table~\ref{tab:gravandwpparticipants} summarises the 
 channels and signal regions combined in the analysis for the EGM $W'$ and bulk \grav. 
 
 \begin{table}[!pht]
\caption{
  Channels and signal regions contributing to the combination for the EGM $W'$ and bulk \grav.
\label{tab:gravandwpparticipants}
}
\begin{center}
\begin{tabular}{c|c|c|c}
\hline
\hline
Channel & Signal region &   $W'$ mass range [\TeV] &   \grav{} mass range [\TeV] \\
\hline
\multirow{2}{*}{\lnull} & low-mass & 0.2--1.9 &  -- \\
& high-mass & 0.2--2.5 & -- \\
\hline
\multirow{3}{*}{\llqq} & low-\pt{} resolved & 0.3--0.9 & 0.2--0.9  \\
& high-\pt{} resolved & 0.6--2.5 &  0.6--0.9\\
& merged  & 0.9--2.5 &  0.9--2.5 \\ 
\hline
\multirow{3}{*}{\lnuqq} & low-\pt{} resolved & 0.3--0.8 & 0.2--0.7  \\
& high-\pt{} resolved & 0.6--1.1 & 0.6--0.9\\
& merged  & 0.8--2.5 & 0.8--2.5\\ 
\hline
\multirow{3}{*}{$JJ$}  & $WZ$ selection &  1.3--2.5 & -- \\
& $WW$+$ZZ$ selection  & -- & 1.3--2.5 \\
\hline
\hline
\end{tabular}
\end{center}
\end{table}

%% file: systematics.tex
\section{Systematic uncertainties}
\label{sec:syst}
The sources of systematic uncertainty along with their effects on the expected signal and background yields for each of the individual channels used in this combination are described in detail in their corresponding publications~\cite{Aad:2014pha,Aad:2014xka,Aad:2015ufa,Aad:2015owa}. Although the results from the different search channels in this combination are statistically independent, commonalities between the different search channels, such as the objects used, the signal and background simulation, and the integrated luminosity estimation, introduce correlated effects in the signal and background expectations. 
Whenever an effect due to an uncertainty in the triggering, identification, or reconstruction of leptons is considered for a channel, it is treated as fully correlated with the effects due to this uncertainty in other channels.

  In the same manner, the effects of each uncertainty related to the small-$R$ jet energy scale and resolution are treated as fully correlated in all channels using small-$R$ jets or \met{}. 
  For the search channels using large-$R$ jets, uncertainties in the large-$R$ jet energy scale, energy resolution, mass scale, mass resolution, or in the modelling of the boson-tagging discriminant $\sqrt{y}$ are taken as fully correlated.  Uncertainties in the data-driven background estimates are treated as uncorrelated. The effects of uncertainty in the initial- and final-state radiation (ISR and FSR) modelling and in the PDFs are each treated as fully correlated across all search channels. 
   
The effect of a single source of systematic uncertainty on the combined limit can be ranked 
by the loss in sensitivity caused by its inclusion. To quantify the loss of sensitivity 
at a given mass point the value computed with all systematic 
uncertainties included is compared to the value obtained excluding the 
single systematic uncertainty. 
In the low mass region at $\lesssim0.5$ \tev{} the leading uncertainty is the modelling of the SM diboson 
background in the dominant \lnull{} channel with an impact of 35\% sensitivity
degradation in the combined limit for EGM $W'$. The leading source of uncertainty in case of the \grav{} limit
is the modelling of the $\Zzero + $jets background in the \lnuqq channel with a  
degradation of 25\%. In the intermediate mass region 
up to $\lesssim1.5$ \tev{} the uncertainty on the normalisation of the $W$+jets
background in the \lnuqq{} channel is dominating with 20\% to 30\% degradation of the EGM $W'$ limit
and 25\% to 55\% degradation of the \grav{} limit depending
on the mass point, while in the high mass region up to $2$ \tev{} the shape 
uncertainty on the $W$+jets background dominates with a degradation of around 25\% for 
the EGM $W'$ limit and 35\% for the \grav{} limit.

%% file: results.tex
\section{Results}
\label{sec:results}

\begin{figure}[th!]
  \begin{center}
     \subfigure[]
              {
     \includegraphics[width=0.485\textwidth]{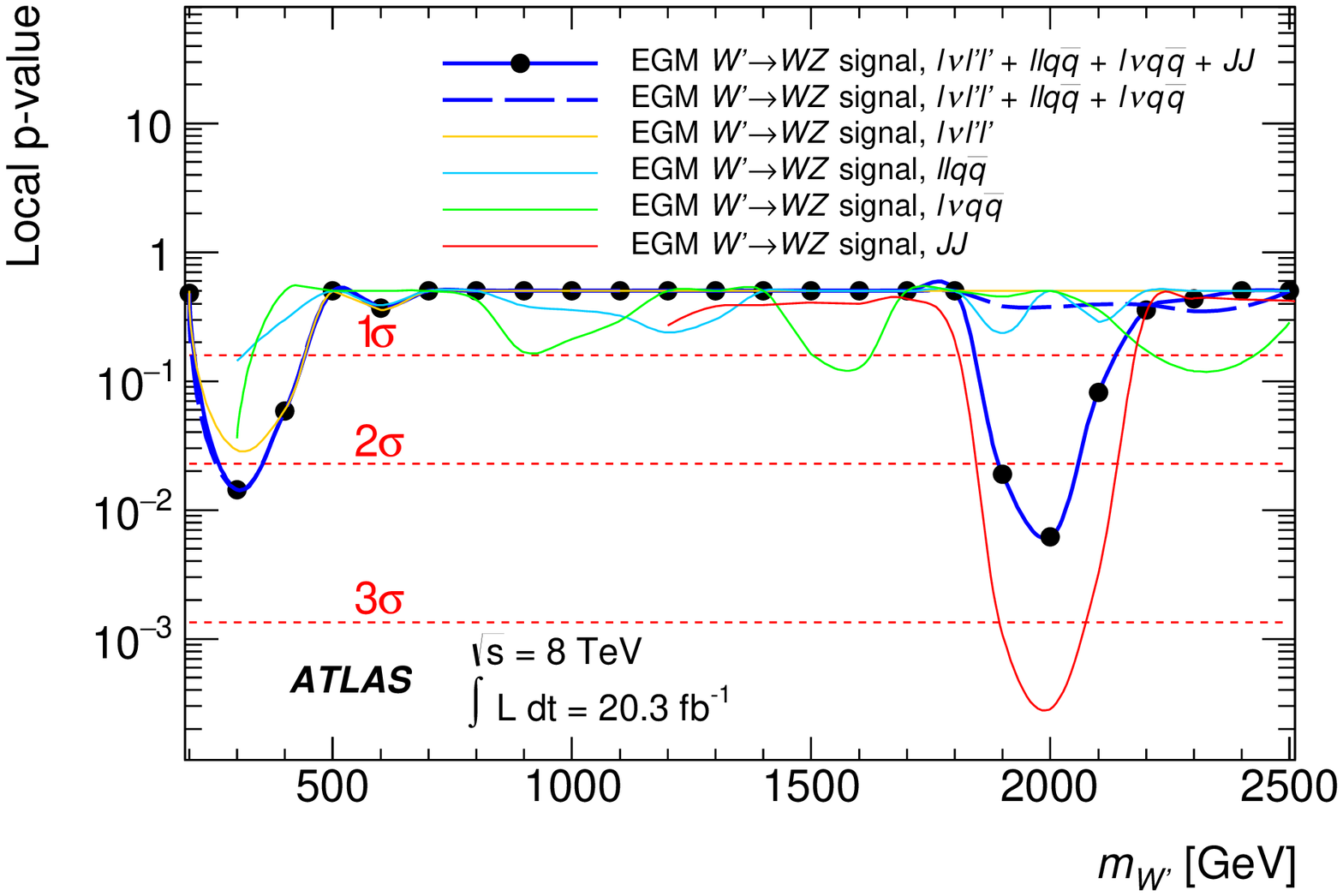}\label{fig:Wpm_pvalue}
              }
    \subfigure[]
              {
     \includegraphics[width=0.485\textwidth]{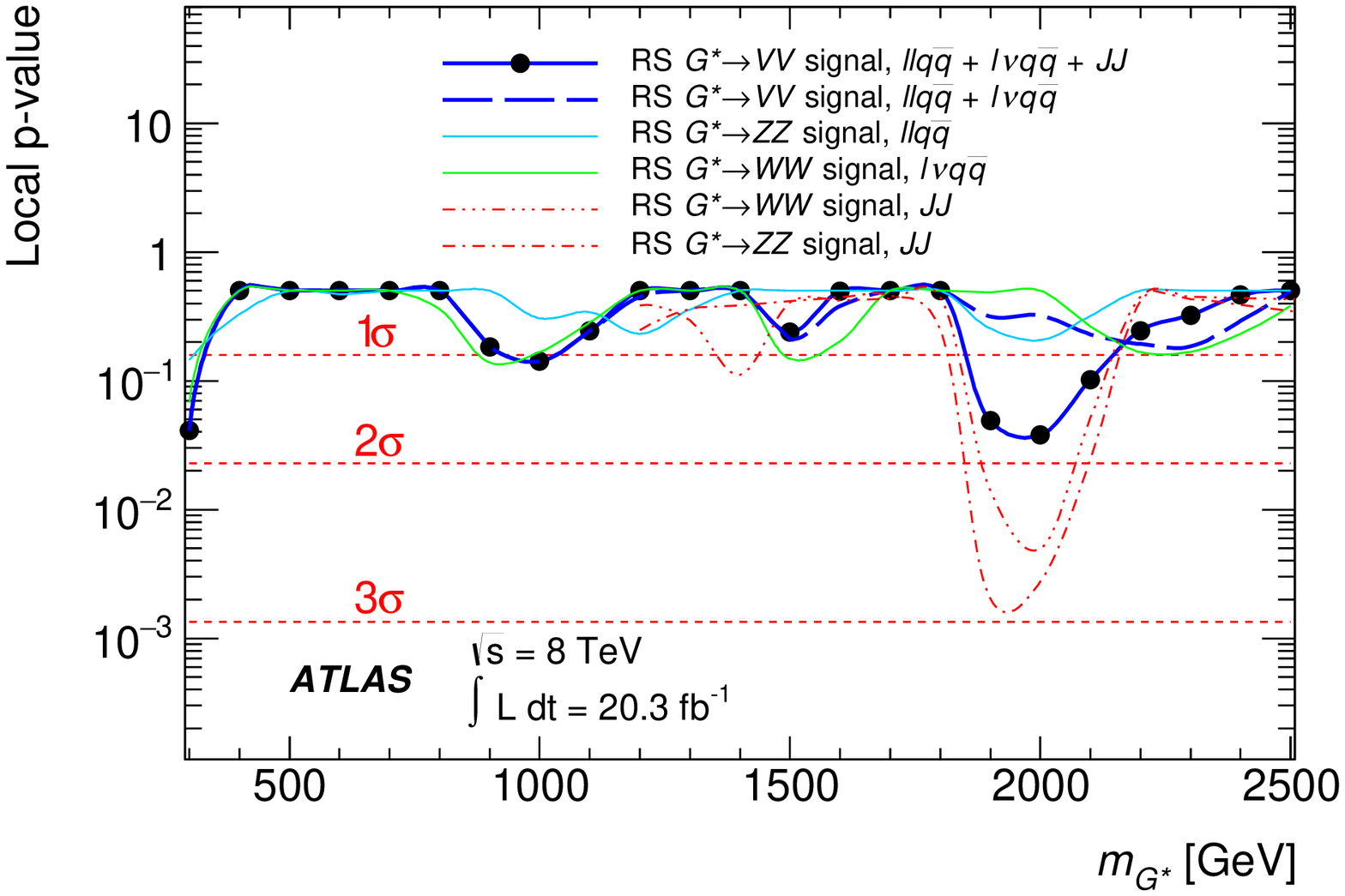}\label{fig:grav_pvalue}
              }
   \end{center}
\vspace{-20 pt}
     \caption{ The $p_0$-value for the individual and combined channels for 
     (a) the EGM $W'$ search in the \lnull, \llqq, \lnuqq{} and $JJ$ channels and 
     (b) the bulk \grav{} search in the \llqq, \lnuqq{} and $JJ$ channels.
     }
\label{fig:pvalues}
\end{figure}

Figure~\ref{fig:pvalues} shows the $p_0$-value obtained in the search for the 
EGM $W'$ and \grav{} as a function of the resonance mass for the  
\lnull, \llqq, \lnuqq{} and $JJ$ channels combined and for the individual channels. 
For the full combination the largest deviation 
from the background-only expectation is found in the EGM $W'$ search at around 2.0 \TeV{} with a $p_0$-value 
corresponding to $2.5$ standard deviations ($\sigma$). This is smaller than the $p_0$-value of $3.4 \,\sigma$ observed 
in the $JJ$ channel alone because the \lnull, \llqq, and \lnuqq{} channels are more 
consistent with the background-only hypothesis. 

The compatibility of the individual channels is quantified with the test described 
in Section~\ref{sec:combproc}. 
In the mass region around 2 \tev{} the $JJ$ channel presents an
excess while the other channels are in good agreement with the background-only expectation.
For the EGM $W'$ benchmark the compatibility of the combined \lnull, \llqq, and \lnuqq{} channels 
with the $JJ$ channel is at the level of $2.9\,\sigma$. 
When accounting for the probability for any of the four channels to fluctuate the compatibility 
is found to be at the level of $2.6 \, \sigma$. 
In comparison the corresponding test for the bulk \grav{} interpretation shows better compatibility.

\begin{figure}[th!]
  \begin{center}
     \subfigure[]
              {
     \includegraphics[width=0.485\textwidth]{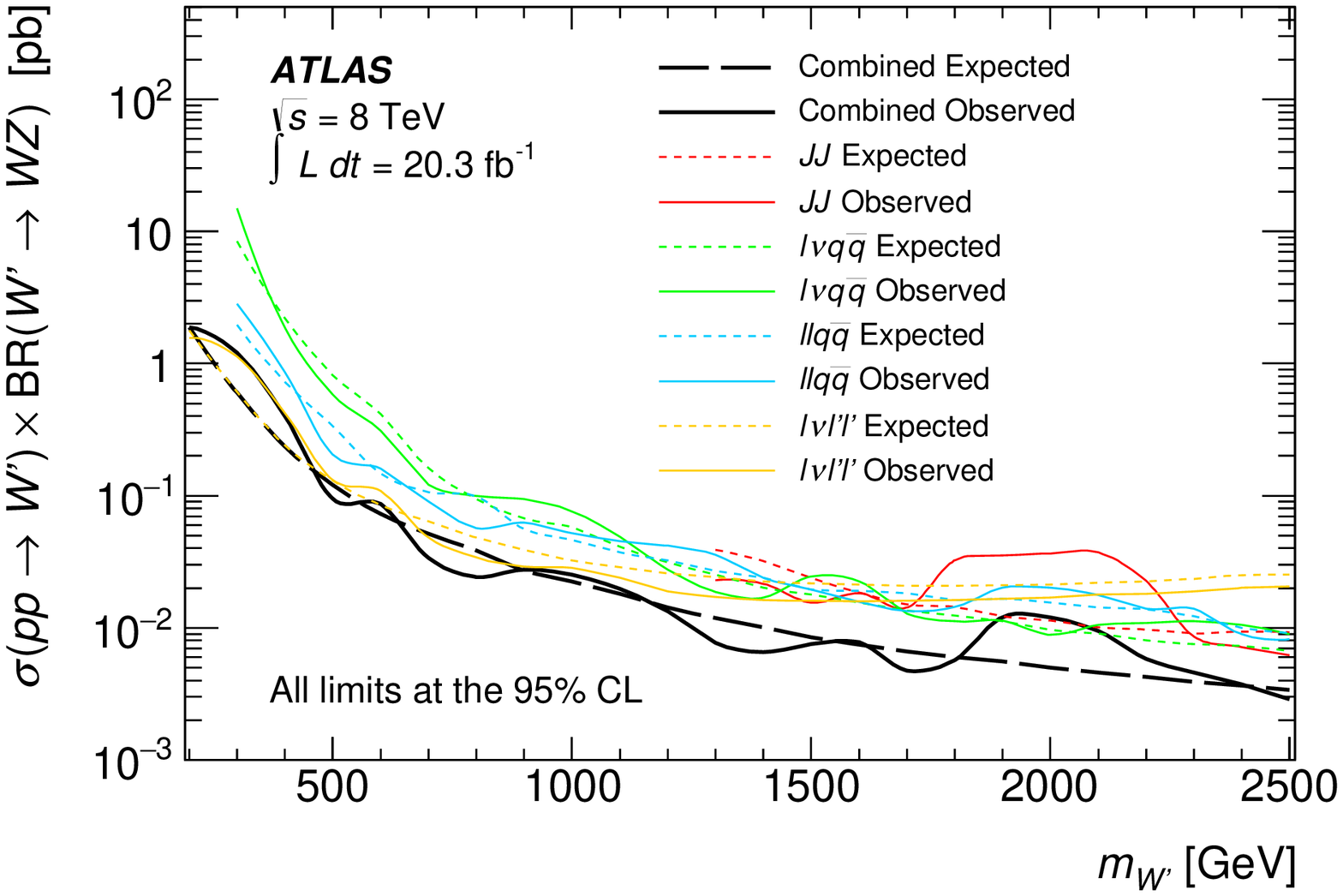}\label{fig:Wpm_all_statandsyst_each}
              }
    \subfigure[]
              {
     \includegraphics[width=0.485\textwidth]{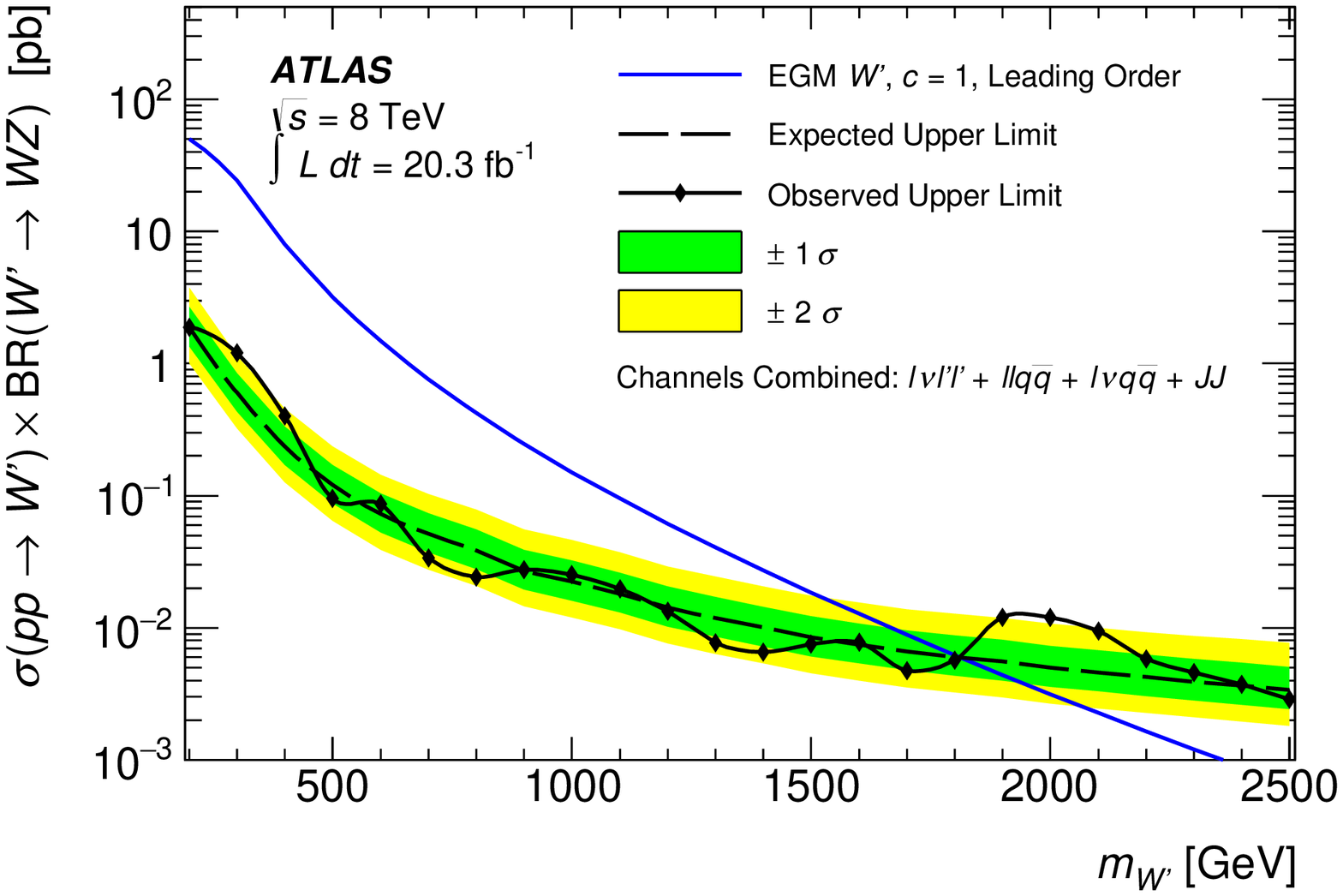}\label{fig:Wpm_all_statandsyst_comb}
              }
   \end{center}
\vspace{-20 pt}
     \caption{The 95\% CL limits on (a) the EGM $W'$ using the 
     \lnull, \llqq, \lnuqq, and $JJ$ channels and their combination, and 
     (b) the combined 95\% CL limit with the green (yellow) bands representing
     the $1 \, \sigma$ ($2 \, \sigma$) intervals of the expected limit 
     including statistical and systematic uncertainties.
}
\label{fig:Wpm_all_statandsyst}
\end{figure}

Figure~\ref{fig:Wpm_all_statandsyst} shows the combined upper limit on the EGM $W'$ production cross 
section times its branching ratio to $\Wboson\Zzero$ at the 95\% CL in the mass range 
from $300\GeV$ to $2.5\TeV$. 
In Fig.~\ref{fig:Wpm_all_statandsyst_each} the observed and expected limits
of the individual and combined channels are shown. 
In Fig.~\ref{fig:Wpm_all_statandsyst_comb} the observed and expected combined
limits are compared with the theoretical EGM $W'$ prediction. 
The resulting combined lower limit on the EGM $W'$ mass 

using a LO cross-section calculation
is observed to be 1.81~\TeV, with an expected limit of 1.81~\TeV.
The most stringent observed mass limit from an individual channel
is $1.59\TeV$ at NNLO in the \lnuqq{} analysis.

\begin{figure}[th!]
  \begin{center}
     \subfigure[]
              {
     \includegraphics[width=0.485\textwidth]{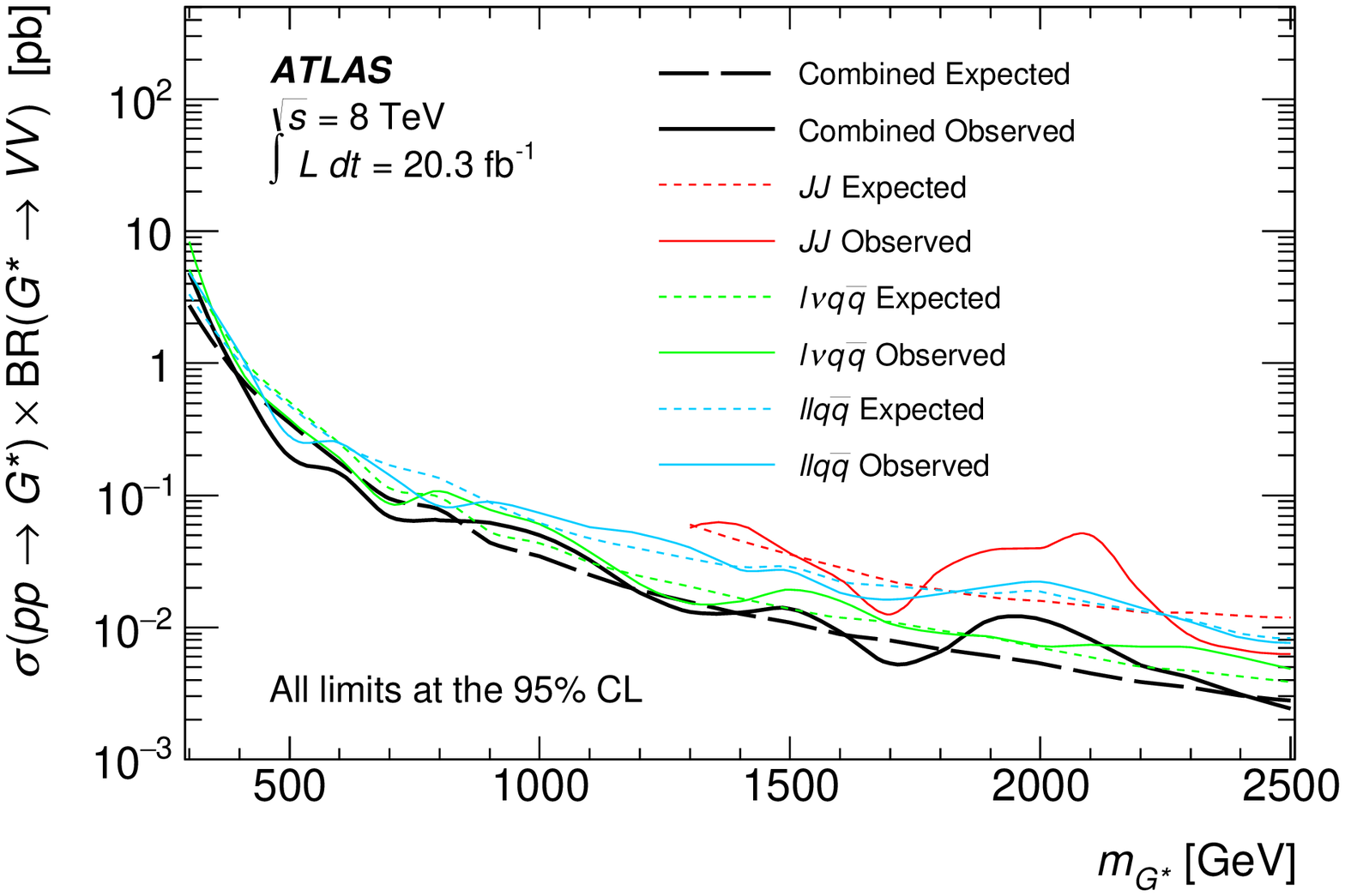}\label{fig:grav_all_statandsyst_each}
              }
    \subfigure[]
              {
     \includegraphics[width=0.485\textwidth]{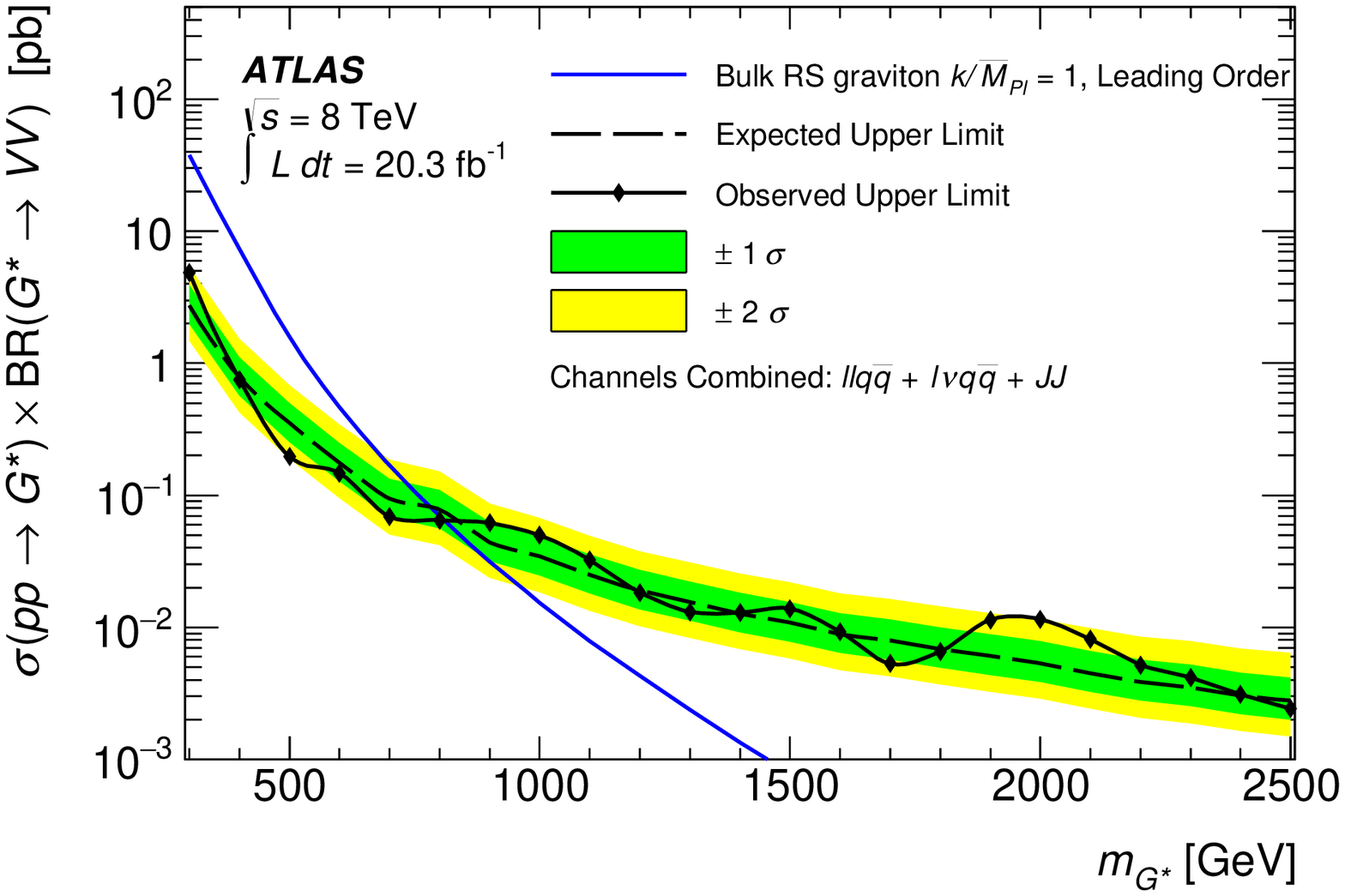}\label{fig:grav_all_statandsyst_comb}
              }
   \end{center}
\vspace{-20 pt}
     \caption{
     The 95\% CL limits on (a) the bulk \grav{} using the \llqq, \lnuqq, 
     and $JJ$ channels and their combination, and (b) the combined 95\% CL limit with the 
     green (yellow) bands representing the $1 \, \sigma$ ($2 \, \sigma$) intervals of the expected limit 
     including statistical and systematic uncertainties.
    }
\label{fig:grav_all_statandsyst}
\end{figure}

In Fig.~\ref{fig:grav_all_statandsyst} the observed and expected upper limits at the 95\% CL 
on the bulk $\grav$ production cross section times its branching ratio 
to $\Wboson\Wboson$ and $\Zzero\Zzero$ are shown in the mass range from $200\GeV$ to $2.5\TeV$.
In Fig.~\ref{fig:grav_all_statandsyst_comb} the observed and expected limits
of the individual and combined channels are shown and compared with the theoretical 
bulk \grav{} prediction for $k/\bar{M}_{\mathrm{Pl}} = 1$.
The combined, lower mass limit for the bulk $\grav$, assuming $k/\bar{M}_{\mathrm{Pl}} = 1$,  
is 810~\GeV, with an expected limit of 790~\GeV. 
The most stringent lower mass limit from the individual \llqq, \lnuqq{} and $JJ$ channels
is $760\GeV$ from the \lnuqq{} channel.

\clearpage

%% file: conclusion.tex
\section{Conclusion}
\label{sec:conclusion}
A combination of individual searches in all-leptonic, semileptonic, 
and all-hadronic final states to search for new heavy bosons decaying 
to $\Wboson\Wboson$, $\Wboson\Zzero$ and $\Zzero\Zzero$ is presented.
The searches use 20.3 fb$^{-1}$ of 8 \TeV{} $pp$ collision data collected by the ATLAS detector at the LHC.
Within the combined result, no significant excess over the background-only expectation in the invariant mass distribution of the diboson candidates is observed.
Upper limits on the production cross section times branching ratio to dibosons at the 
95\% CL are evaluated within the context of an extended gauge model with a heavy $W'$ 
boson and a bulk Randall--Sundrum model with a heavy spin-2 graviton.
The combination significantly improves both the cross-section limits and the mass limits for 
EGM $W'$ and bulk $\grav$ production over the most stringent limits
of the individual analyses. The observed lower limit on the EGM $W'$ mass 
is found to be 1.81~\TeV{} and for the bulk $\grav$ mass, assuming $k/\bar{M}_{\mathrm{Pl}} = 1$, the observed limit is 810~\GeV. 

%% file: acknowledgements/Acknowledgements.tex
% Acknowledgements for papers with collision data
% Version 15-Sep-2015

% Standard acknowledgements start here
%----------------------------------------------
We thank CERN for the very successful operation of the LHC, as well as the
support staff from our institutions without whom ATLAS could not be
operated efficiently.

We acknowledge the support of ANPCyT, Argentina; YerPhI, Armenia; ARC, Australia; BMWFW and FWF, Austria; ANAS, Azerbaijan; SSTC, Belarus; CNPq and FAPESP, Brazil; NSERC, NRC and CFI, Canada; CERN; CONICYT, Chile; CAS, MOST and NSFC, China; COLCIENCIAS, Colombia; MSMT CR, MPO CR and VSC CR, Czech Republic; DNRF, DNSRC and Lundbeck Foundation, Denmark; IN2P3-CNRS, CEA-DSM/IRFU, France; GNSF, Georgia; BMBF, HGF, and MPG, Germany; GSRT, Greece; RGC, Hong Kong SAR, China; ISF, I-CORE and Benoziyo Center, Israel; INFN, Italy; MEXT and JSPS, Japan; CNRST, Morocco; FOM and NWO, Netherlands; RCN, Norway; MNiSW and NCN, Poland; FCT, Portugal; MNE/IFA, Romania; MES of Russia and NRC KI, Russian Federation; JINR; MESTD, Serbia; MSSR, Slovakia; ARRS and MIZ\v{S}, Slovenia; DST/NRF, South Africa; MINECO, Spain; SRC and Wallenberg Foundation, Sweden; SERI, SNSF and Cantons of Bern and Geneva, Switzerland; MOST, Taiwan; TAEK, Turkey; STFC, United Kingdom; DOE and NSF, United States of America. In addition, individual groups and members have received support from BCKDF, the Canada Council, CANARIE, CRC, Compute Canada, FQRNT, and the Ontario Innovation Trust, Canada; EPLANET, ERC, FP7, Horizon 2020 and Marie Skłodowska-Curie Actions, European Union; Investissements d'Avenir Labex and Idex, ANR, Region Auvergne and Fondation Partager le Savoir, France; DFG and AvH Foundation, Germany; Herakleitos, Thales and Aristeia programmes co-financed by EU-ESF and the Greek NSRF; BSF, GIF and Minerva, Israel; BRF, Norway; the Royal Society and Leverhulme Trust, United Kingdom.

The crucial computing support from all WLCG partners is acknowledged
gratefully, in particular from CERN and the ATLAS Tier-1 facilities at
TRIUMF (Canada), NDGF (Denmark, Norway, Sweden), CC-IN2P3 (France),
KIT/GridKA (Germany), INFN-CNAF (Italy), NL-T1 (Netherlands), PIC (Spain),
ASGC (Taiwan), RAL (UK) and BNL (USA) and in the Tier-2 facilities
worldwide.
%----------------------------------------------

%% file: atlas_authlist.tex
% ATLAS Collaboration author list
% Data extracted on 21-Sep-2015 for paper reference EXOT-2014-18
%\documentclass[11pt]{article}
%\usepackage{a4wide}\begin{document}
\begin{flushleft}
{\Large The ATLAS Collaboration}

\bigskip

G.~Aad$^{\rm 85}$,
B.~Abbott$^{\rm 113}$,
J.~Abdallah$^{\rm 151}$,
O.~Abdinov$^{\rm 11}$,
R.~Aben$^{\rm 107}$,
M.~Abolins$^{\rm 90}$,
O.S.~AbouZeid$^{\rm 158}$,
H.~Abramowicz$^{\rm 153}$,
H.~Abreu$^{\rm 152}$,
R.~Abreu$^{\rm 116}$
% ...
\bigskip
\\
% ...
$^{11}$ Institute of Physics, Azerbaijan Academy of Sciences, Baku, Azerbaijan\\
% ...
$^{85}$ CPPM, Aix-Marseille Universit{\'e} and CNRS/IN2P3, Marseille, France\\
% ...
$^{90}$ Department of Physics and Astronomy, Michigan State University, East Lansing MI, United States of America\\
% ...
$^{107}$ Nikhef National Institute for Subatomic Physics and University of Amsterdam, Amsterdam, Netherlands\\
% ...
$^{113}$ Homer L. Dodge Department of Physics and Astronomy, University of Oklahoma, Norman OK, United States of America\\
% ...
$^{116}$ Center for High Energy Physics, University of Oregon, Eugene OR, United States of America\\
% ...
$^{151}$ Institute of Physics, Academia Sinica, Taipei, Taiwan\\
$^{152}$ Department of Physics, Technion: Israel Institute of Technology, Haifa, Israel\\
$^{153}$ Raymond and Beverly Sackler School of Physics and Astronomy, Tel Aviv University, Tel Aviv, Israel\\
% ...
$^{158}$ Department of Physics, University of Toronto, Toronto ON, Canada\\
% ...
\end{flushleft}